\documentclass[12pt]{article}

\usepackage{scicite}

\usepackage{times}
\usepackage{graphicx}
\usepackage{amssymb}

\newenvironment{sciabstract}{%
\begin{quote} \bf}
{\end{quote}}

\newcounter{lastnote}
\newenvironment{scilastnote}{%
\setcounter{lastnote}{\value{enumiv}}%
\addtocounter{lastnote}{+1}%
\begin{list}%
{\setlength{\leftmargin}{.22in}}
{\setlength{\labelsep}{.5em}}}
{\end{list}}
\newcommand{\etal}{\MakeLowercase{\textit{et al.}}} 

\title{Detection of the Characteristic Pion-decay Signature in Supernova Remnants}  

\date{}
\begin{document}
\pagestyle{empty}
\maketitle

\noindent
M.~Ackermann$^{1}$, 
M.~Ajello$^{2}$, 
A.~Allafort$^{3}$, 
L.~Baldini$^{4}$, 
J.~Ballet$^{5}$, 
G.~Barbiellini$^{6,7}$, 
M.~G.~Baring$^{8}$, 
D.~Bastieri$^{9,10}$, 
K.~Bechtol$^{3}$, 
R.~Bellazzini$^{11}$
% ...
\medskip
\begin{enumerate}
\item[1.] Deutsches Elektronen Synchrotron DESY, D-15738 Zeuthen, Germany
\item[2.] Space Sciences Laboratory, 7 Gauss Way, University of California, Berkeley, CA 94720-7450, USA
\item[3.] W. W. Hansen Experimental Physics Laboratory, Kavli Institute for Particle Astrophysics and Cosmology, Department of Physics and SLAC National Accelerator Laboratory, Stanford University, Stanford, CA 94305, USA
\item[4.] Universit\`a  di Pisa and Istituto Nazionale di Fisica Nucleare, Sezione di Pisa I-56127 Pisa, Italy
\item[5.] Laboratoire AIM, CEA-IRFU/CNRS/Universit\'e Paris Diderot, Service d'Astrophysique, CEA Saclay, 91191 Gif sur Yvette, France
\item[6.] Istituto Nazionale di Fisica Nucleare, Sezione di Trieste, I-34127 Trieste, Italy
\item[7.] Dipartimento di Fisica, Universit\`a di Trieste, I-34127 Trieste, Italy
\item[8.] Rice University, Department of Physics and Astronomy, MS-108, P. O. Box 1892, Houston, TX 77251, USA
\item[9.] Istituto Nazionale di Fisica Nucleare, Sezione di Padova, I-35131 Padova, Italy
\item[10.] Dipartimento di Fisica e Astronomia "G. Galilei", Universit\`a di Padova, I-35131 Padova, Italy
\item[11.] Istituto Nazionale di Fisica Nucleare, Sezione di Pisa, I-56127 Pisa, Italy
\item[12.] Istituto Nazionale di Fisica Nucleare, Sezione di Perugia, I-06123 Perugia, Italy
\item[13.] Dipartimento di Fisica, Universit\`a degli Studi di Perugia, I-06123 Perugia, Italy
\item[14.] NASA Goddard Space Flight Center, Greenbelt, MD 20771, USA
\item[15.] Dipartimento di Fisica ``M. Merlin" dell'Universit\`a e del Politecnico di Bari, I-70126 Bari, Italy
\item[16.] Istituto Nazionale di Fisica Nucleare, Sezione di Bari, 70126 Bari, Italy
\item[17.] Laboratoire Leprince-Ringuet, \'Ecole polytechnique,
  CNRS/IN2P3, 91128 Palaiseau, France
\item[18.] Institut de Ci\`encies de l'Espai (IEEE-CSIC), Campus UAB, 08193 Barcelona, Spain
\item[19.] INAF-Istituto di Astrofisica Spaziale e Fisica Cosmica, I-20133 Milano, Italy
\item[20.] Center for Research and Exploration in Space Science and Technology (CRESST) and NASA Goddard Space Flight Center, Greenbelt, MD 20771, USA
\item[21.] Department of Physics and Center for Space Sciences and Technology, University of Maryland Baltimore County, Baltimore, MD 21250, USA
\item[22.] Center for Earth Observing and Space Research, College of Science, George Mason University, Fairfax, VA 22030, resident at Naval Research Laboratory, Washington, DC 20375, USA
\item[23.] National Research Council Research Associate, National Academy of Sciences, Washington, DC 20001, resident at Naval Research Laboratory, Washington, DC 20375, USA
\item[24.] INFN and Dipartimento di Fisica e Astronomia "G. Galilei", Universit\`a di Padova, I-35131 Padova, Italy, 
\item[25.] Instituto de Astronom\'ia y Fisica del Espacio, Pabell\'on
  IAFE, Cdad. Universitaria, C1428ZAA, Buenos Aires, Argentina
\item[26.] ASI Science Data Center, I-00044 Frascati (Roma), Italy
\item[27.] Laboratoire Univers et Particules de Montpellier, Universit\'e Montpellier 2, CNRS/IN2P3, Montpellier, France
\item[28.] Department of Physics and Astronomy, Sonoma State University, Rohnert Park, CA 94928-3609, USA
\item[29.] Department of Physics, Stockholm University, AlbaNova, SE-106 91 Stockholm, Sweden
\item[30.] The Oskar Klein Centre for Cosmoparticle Physics, AlbaNova, SE-106 91 Stockholm, Sweden
\item[31.] Royal Swedish Academy of Sciences Research Fellow, funded by a grant from the K. A. Wallenberg Foundation
\item[32.] Institut universitaire de France, 75005 Paris, France
\item[33.] Agenzia Spaziale Italiana (ASI) Science Data Center, I-00044 Frascati (Roma), Italy
\item[34.] IASF Palermo, 90146 Palermo, Italy
\item[35.] INAF-Istituto di Astrofisica Spaziale e Fisica Cosmica, I-00133 Roma, Italy
\item[36.] Dipartimento di Fisica, Universit\`a di Udine and Istituto Nazionale di Fisica Nucleare, Sezione di Trieste, Gruppo Collegato di Udine, I-33100 Udine, Italy
\item[37.] Space Science Division, Naval Research Laboratory, Washington, DC 20375-5352, USA
\item[38.] Department of Physical Sciences, Hiroshima University, Higashi-Hiroshima, Hiroshima 739-8526, Japan
\item[39.] INAF Istituto di Radioastronomia, 40129 Bologna, Italy
\item[40.] Max-Planck-Institut f\"ur Kernphysik, D-69029 Heidelberg, Germany
\item[41.] Landessternwarte, Universit\"at Heidelberg, K\"onigstuhl, D 69117 Heidelberg, Germany
\item[42.] Department of Astronomy, Graduate School of Science, Kyoto University, Sakyo-ku, Kyoto 606-8502, Japan
\item[43.] School of Physics and Astronomy, University of Southampton, Highfield, Southampton, SO17 1BJ, UK
\item[44.] Department of Physics, Center for Cosmology and Astro-Particle Physics, The Ohio State University, Columbus, OH 43210, USA
\item[45.] Department of Physics, Royal Institute of Technology (KTH), AlbaNova, SE-106 91 Stockholm, Sweden
\item[46.] Science Institute, University of Iceland, IS-107 Reykjavik, Iceland
\item[47.] Research Institute for Science and Engineering, Waseda University, 3-4-1, Okubo, Shinjuku, Tokyo 169-8555, Japan
\item[48.] CNRS, IRAP, F-31028 Toulouse cedex 4, France
\item[49.] GAHEC, Universit\'e de Toulouse, UPS-OMP, IRAP, 31028 Toulouse, France
\item[50.] Department of Astronomy, Stockholm University, SE-106 91 Stockholm, Sweden
\item[51.] Istituto Nazionale di Fisica Nucleare, Sezione di Torino, I-10125 Torino, Italy
\item[52.] Universit\'e Bordeaux 1, CNRS/IN2p3, Centre d'\'Etudes Nucl\'eaires de Bordeaux Gradignan, 33175 Gradignan, France
\item[53.] Funded by contract ERC-StG-259391 from the European Community
\item[54.] Department of Physics and Department of Astronomy, University of Maryland, College Park, MD 20742, USA
\item[55.] Mullard Space Science Laboratory, University College London, Holmbury St. Mary, Dorking, Surrey, RH5 6NT, UK
\item[56.] Hiroshima Astrophysical Science Center, Hiroshima University, Higashi-Hiroshima, Hiroshima 739-8526, Japan
\item[57.] Istituto Nazionale di Fisica Nucleare, Sezione di Roma ``Tor Vergata", I-00133 Roma, Italy
\item[58.] Institute of Space and Astronautical Science, JAXA, 3-1-1 Yoshinodai, Chuo-ku, Sagamihara, Kanagawa 252-5210, Japan
\item[59.] Department of Physics and Astronomy, University of Denver, Denver, CO 80208, USA
\item[60.] Max-Planck-Institut f\"ur Physik, D-80805 M\"unchen, Germany
\item[61.] Harvard-Smithsonian Center for Astrophysics, Cambridge, MA 02138, USA
\item[62.] Santa Cruz Institute for Particle Physics, Department of Physics and Department of Astronomy and Astrophysics, University of California at Santa Cruz, Santa Cruz, CA 95064, USA
\item[63.] Institut f\"ur Astro- und Teilchenphysik and Institut f\"ur Theoretische Physik, Leopold-Franzens-Universit\"at Innsbruck, A-6020 Innsbruck, Austria
\item[64.] NYCB Real-Time Computing Inc., Lattingtown, NY 11560-1025, USA
\item[65.] Department of Physics and Astronomy, University of California, Los Angeles, CA 90095-1547, USA
\item[66.] Max-Planck Institut f\"ur extraterrestrische Physik, 85748 Garching, Germany
\item[67.] Department of Chemistry and Physics, Purdue University Calumet, Hammond, IN 46323-2094, USA
\item[68.] Solar-Terrestrial Environment Laboratory, Nagoya University, Nagoya 464-8601, Japan
\item[69.] Department of Physics, Graduate School of Science, Kyoto
  University, Sakyo-ku, Kyoto 606-8502, Japan
\item[70.] Department of Physics, Willamette University, Salem, OR 97031, USA
\item[71.] Institut f\"ur Theoretische Physik and Astrophysik,
  Universit\"at W\"urzburg, D-97074 W\"urzburg, Germany 
\item[72.] NASA Postdoctoral Program Fellow
\item[73.] Consorzio Interuniversitario per la Fisica Spaziale, I-10133 Torino, Italy
\item[74.] Dipartimento di Fisica, Universit\`a di Roma ``Tor Vergata", I-00133 Roma, Italy
\item[75.] Department of Physics and Mathematics, Aoyama Gakuin University, Sagamihara, Kanagawa, 252-5258, Japan
\item[$\star$] funk@slac.stanford.edu, $\dagger$
  ttanaka@cr.scphys.kyoto-u.ac.jp, $\ddagger$ uchiyama@slac.stanford.edu
\end{enumerate}

\begin{sciabstract} {\bf{Cosmic rays are particles (mostly protons)
      accelerated to relativistic speeds. Despite wide agreement that
      supernova remnants (SNRs) are the sources of galactic cosmic
      rays, unequivocal evidence for the acceleration of protons in
      these objects is still lacking. When accelerated protons
      encounter interstellar material they produce neutral pions,
      which in turn decay into gamma rays. This offers a compelling
      way to detect the acceleration sites of protons. The
      identification of pion-decay gamma rays has been difficult
      because high-energy electrons also produce gamma rays via
      bremsstrahlung and inverse Compton scattering. We detected the
      characteristic pion-decay feature in the gamma-ray spectra of
      two SNRs, IC 443 and W44, with the Fermi Large Area
      Telescope. This detection provides direct evidence that
      cosmic-ray protons are accelerated in SNRs.}}
\end{sciabstract}

A supernova explosion drives its progenitor material supersonically
into interstellar space, forming a collisionless shock wave ahead of
the stellar ejecta.  The huge amount of kinetic energy released by a
supernova, typically $10^{51}\ \rm ergs$, is initially carried by the
expanding ejecta and is then transferred to kinetic and thermal
energies of shocked interstellar gas and relativistic particles. The
shocked gas and relativistic particles produce the thermal and
nonthermal emissions of a supernova remnant (SNR).  The mechanism of
diffusive shock acceleration (DSA) can explain the production of
relativistic particles in SNRs \cite{MalkovDrury}.  DSA generally
predicts that a substantial fraction of the shock energy is
transferred to relativistic protons.  Indeed, if SNRs are the main
sites of acceleration of the galactic cosmic rays, then 3 to 30\% of
the supernova kinetic energy must end up transferred to relativistic
protons. However, the presence of relativistic protons in SNRs has
been mostly inferred from indirect
arguments\cite{HughesContactDiscontinuity,UchiyamaVariability,HelderLineWidth,TychoMorlino}.

A direct signature of high energy protons is provided by gamma rays
generated in the decay of neutral pions ($\pi^0$); proton-proton (more
generally nuclear-nuclear) collisions create $\pi^0$ mesons which
usually quickly decay into two gamma rays\cite{Dermer86,DAV94,Naito94}
(schematically written as $p+p \rightarrow \pi^0$+ {\emph{other
    products}}, followed by $\pi^0 \rightarrow 2 \gamma $), each
having an energy of $m_{\pi^0}c^2/2 = 67.5$ MeV in the rest frame of
the neutral pion (where $m_{\pi^0}$ is the rest mass of the neutral
pion and $c$ is the speed of light). The gamma-ray number spectrum,
$F(\varepsilon )$, is thus symmetric about 67.5 MeV in a log-log
representation~\cite{Stecker1970}.  The $\pi^0$-decay spectrum in the
usual $\varepsilon^2 F(\varepsilon )$ representation rises steeply
below $\sim 200$ MeV and approximately traces the energy distribution
of parent protons at energies greater than a few GeV.  This
characteristic spectral feature (often referred to as the ``pion-decay
bump") uniquely identifies $\pi^0$-decay gamma rays and thereby
high-energy protons, allowing a measurement of the source spectrum of
cosmic rays.

Massive stars are short-lived and end their lives with core-collapse
supernova explosions. These explosions typically occur in the vicinity
of molecular clouds with which they interact.  When cosmic-ray protons
accelerated by SNRs penetrate into high density clouds, $\pi^0$-decay
gamma-ray emission is expected to be enhanced because of more frequent
$pp$ interactions relative to the interstellar
medium\cite{AharonianCloud}.  Indeed, SNRs interacting with molecular
clouds are the most luminous SNRs in gamma
rays\cite{Fermi_W51C,ThompsonCR}.  The best examples of SNR-cloud
interactions in our galaxy are the SNRs IC~443 and W44\cite{Seta98},
which are the two highest-significance SNRs in the second
{\emph{Fermi}} Large Area Telescope (LAT) catalog (2FGL)~\cite{2FGL}
and are thus particularly suited for a dedicated study of the details
of their gamma-ray spectra. The age of each remnant is estimated to be
$\sim 10,000$ years.  IC~443 and W44 are located at distances of 1.5 kpc
and 2.9 kpc, respectively.

We report here on 4 years of observations with the \emph{Fermi} LAT (4
August 2008 to 16 July 2012) of IC~443 and W44, focusing on the
sub-GeV part of the gamma-ray spectrum -- a crucial spectral window
for distinguishing $\pi^0$-decay gamma rays from electron
bremsstrahlung or inverse Compton scattering produced by relativistic
electrons. Previous analyses of IC~443 and W44 used only 1 year of
{\emph{Fermi}} LAT data~\cite{Fermi_W44,Fermi_IC443, ExtendedFermi}
and were limited to the energy band above 200 MeV, mainly because of
the small and rapidly changing LAT effective area at low energies. A
recent update to the event classification and background rejection
(so-called {\emph{Pass 7}}) provides an increase in LAT effective area
at 100 MeV by a factor of $\sim 5$~\cite{LATPerformance}, enabling the
study of bright, steady sources in the galactic plane below 200 MeV
with the \emph{Fermi}-LAT. Note that the gamma-ray spectral energy
distribution of W44 measured recently by the AGILE satellite falls
steeply below 1 GeV, which the authors interpreted as a clear
indication for the $\pi^0$-decay origin of the gamma-ray
emission\cite{AGILE_W44}.  Also, a recent analysis of W44 at high
energies (above 2 GeV) has been reported~\cite{UchiyamaW44},
revealing large-scale gamma-ray emission attributable to high-energy
protons that have escaped from W44. Here we present analyses of the
gamma-ray emission from the compact regions delineated by the radio
continuum emission of IC~443 and W44.

\begin{figure}[!t]
\centering
\includegraphics[width=\textwidth]{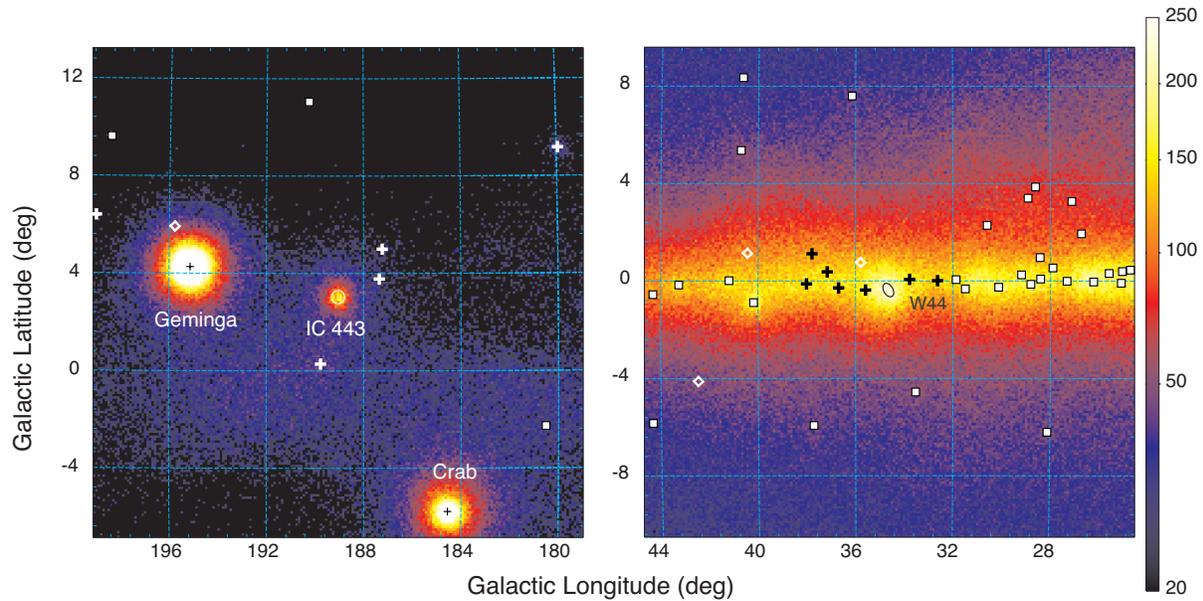} 
\caption{ Gamma-ray count maps of the $20^{\circ}\times20^{\circ}$
  fields around IC~443 (left panel) and W44 (right panel) in the
  energy range 60 MeV to 2 GeV.  Nearby gamma-ray sources are marked
  as crosses and squares. Diamonds denote previously undetected
  sources. For sources indicated by crosses and diamonds, the fluxes
  were left as free parameters in the analysis. Events were spatially
  binned in regions of side length 0.1$^{\circ}$, the units of the
  color bar is the square
  root of count density, and the colors have been clipped at 20
  counts per pixel to make the galactic diffuse emission less
  prominent. Given the spectra of the sources and the effective area
  of the LAT instrument, the bulk of the photons seen in this plot
  have energies between 300 and 500 MeV. IC~443 is located in the
  galactic anti-center region, where the background gamma-ray emission
  produced by the pool of galactic cosmic rays interacting with interstellar
  gas is rather weak relative to the region around W44.  The two
  dominant sources in the IC~443 field are the Geminga pulsar
  (2FGL\,J0633.9+1746) and the Crab (2FGL\,J0534.5+2201). For the W44
  count map, W44 is the dominant source, sub-dominant, however, to the
  galactic diffuse emission.} .
\label{fig::cmap}
\end{figure}

The analysis was performed using the {\emph{Fermi}} LAT {\emph{Science
    Tools}}~\cite{FGSC}.  We used a maximum likelihood technique to
determine the significance of a source over the background and to fit
spectral parameters\cite{Note, Mattox}. For both SNRs, additional
sources seen as excesses in the background-subtracted map have been
added to the background model~\cite{SOM} and are shown as diamonds in
Fig.~\ref{fig::cmap} -- one in the case of IC~443, three in the case
of W44. The inclusion of these sources had no influence on the fitted
spectrum of the SNRs. Three close-by sources (2FGL\,J1852.8+0156c,
2FGL\, J1857.2+0055c, and 2FGL\,J1858.5+0129c) have been identified
with escaping cosmic rays from W44\cite{UchiyamaW44}. These 2FGL
sources have been removed from the background model (see below) in
order to measure the full cosmic-ray content of W44.

\begin{figure}[!t]
\centering
\vspace{-2cm}
\includegraphics[width=0.9\textwidth]{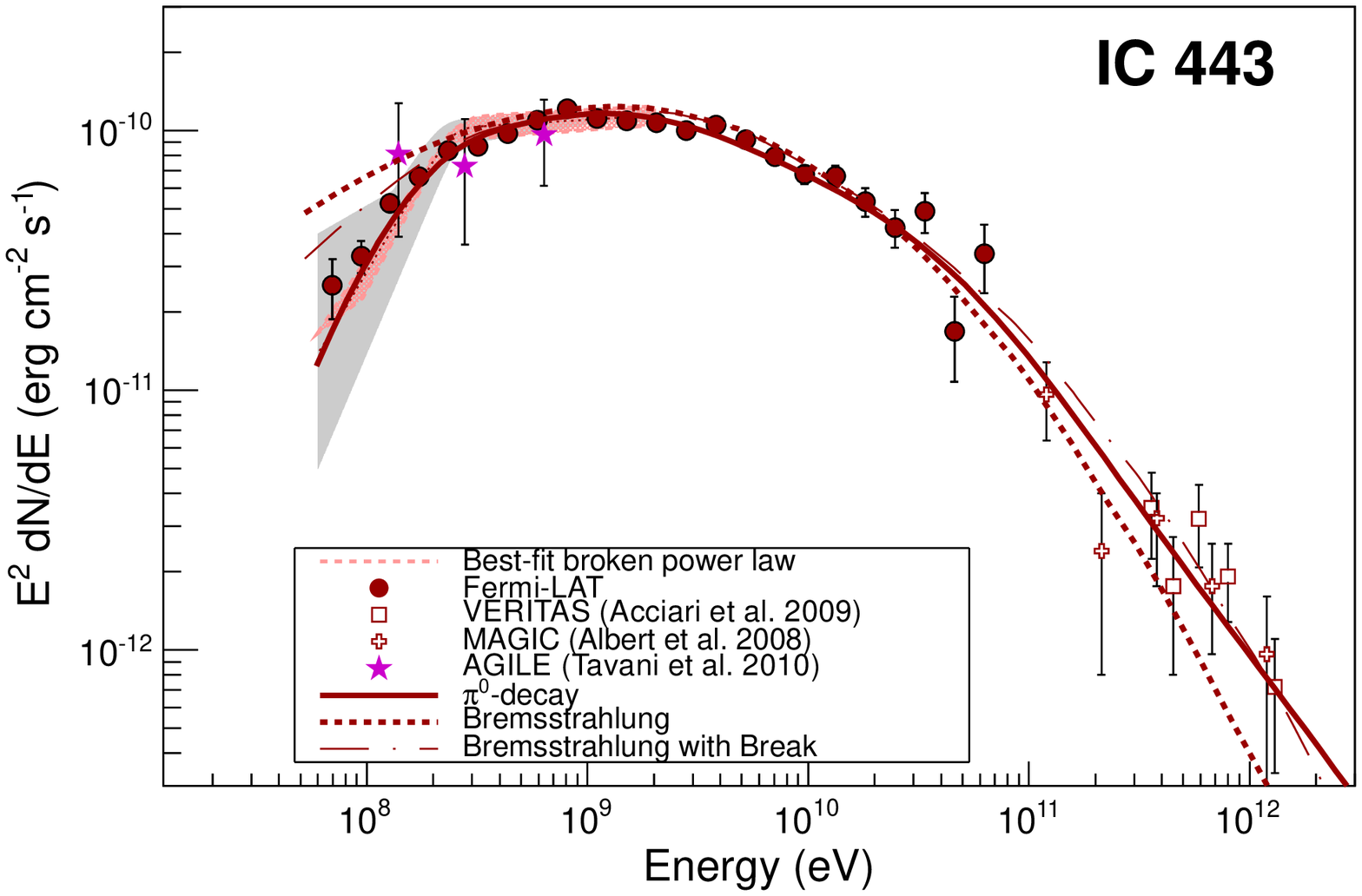}
\includegraphics[width=0.9\textwidth]{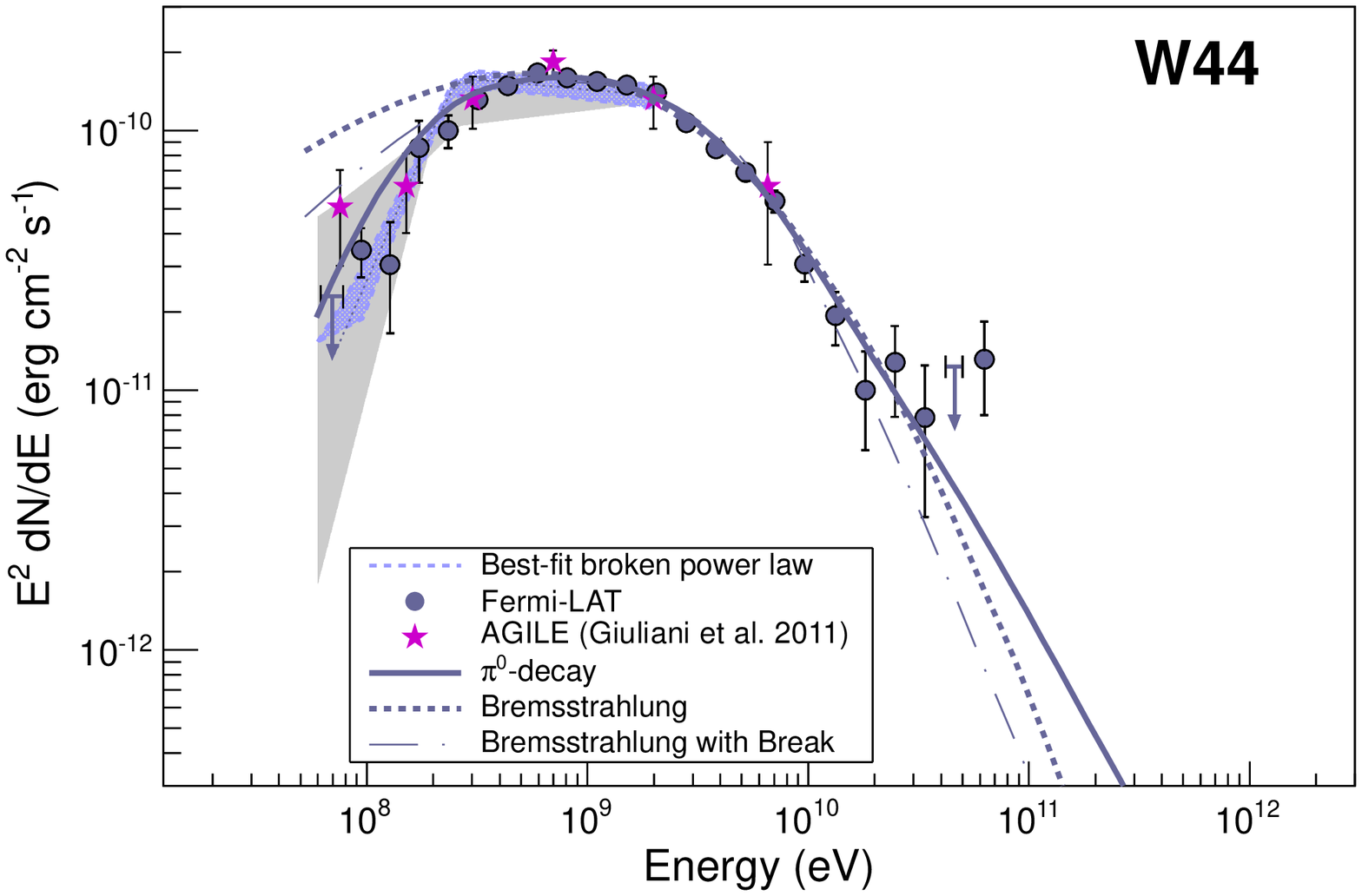}
\caption{ (A and B) Gamma-ray spectra of IC~443 (A) and W44 (B) as measured with the
  {\emph{Fermi}}-LAT. Color-shaded areas bound by dashed lines denote
  the best-fit broadband smooth broken power law (60~MeV to 2~GeV),
  gray-shaded bands show systematic errors below 2 GeV due mainly
  to imperfect modeling of the galactic diffuse emission. At the
  high-energy end, TeV spectral data points for IC~443 from MAGIC
  \cite{MAGIC_IC443} and VERITAS \cite{VERITAS_IC443} are shown.
  Solid lines denote the best-fit pion-decay gamma-ray spectra, dashed
  lines denote the best-fit bremsstrahlung spectra, and dash-dotted lines
  denote the best-fit bremsstrahlung spectra when including an ad hoc
  low-energy break at 300~MeV~$c^{-1}$ in the electron spectrum. These
  fits were done to the {\emph{Fermi}} LAT data alone (not taking the
  TeV data points into account).  Magenta stars denote
  measurements from the AGILE satellite for these two SNRs, taken
  from~\cite{AGILE_IC443} and \cite{AGILE_W44}, respectively.}
\label{fig::spec}
\end{figure}

Figure~\ref{fig::spec} shows the spectral energy distribution obtained
for IC~443 and W44 through maximum likelihood estimation.  To derive
the flux points we performed a maximum likelihood fitting in 24
independent logarithmically spaced energy bands from 60 MeV to 100
GeV. The normalization of the fluxes of IC~443 and W44 and those of
neighboring sources and of the galactic diffuse model, was left free
in the fit for each bin.  In both sources, the spectra below $\sim
200$ MeV are steeply rising, clearly exhibiting a break at $\sim 200$
MeV. To quantify the significances of the spectral breaks, we fit the
fluxes of IC~443 and W44 between 60 MeV and 2 GeV -- below the
high-energy breaks previously found in the 1-year spectra
\cite{Fermi_W44,Fermi_IC443} with both a single power law of the form
$F(\varepsilon ) = K (\varepsilon/\varepsilon_0 )^{-\Gamma_1}$ and a
smoothly broken power law of the form $F(\varepsilon ) = K
(\varepsilon/\varepsilon_0) ^{-\Gamma_1} (1 +
(\varepsilon/\varepsilon_{\rm br})^{(\Gamma_2 -
  \Gamma_1)/\alpha})^{-\alpha}$ with $\varepsilon_0 = 200$ MeV. The
spectral index changes from $\Gamma_1$ to $\Gamma_2$ ($>\Gamma_1$) at
the break energy $\varepsilon_{\rm br}$. The smoothness of the break
is determined by the parameter $\alpha$, which was fixed at $0.1$
(Table~\ref{table}).  We define the test-statistic value ($TS$) as
$2\ln (\mathcal{L}_1 /\mathcal{L}_0 )$ where $\mathcal{L}_{1/0}$
corresponds to the likelihood value for the source/no-source
hypothesis~\cite{Mattox}.  The detection significance is given by
$\sim \sqrt{TS}$.  The smoothly broken power law model yields a
significantly larger $TS$ than a single power law, establishing the
existence of a low-energy break.  The improvement in log likelihood
when comparing the broken power law to a single power law corresponds
to a formal statistical significance of $19 \sigma$ for the low-energy
break in IC~443 and $21 \sigma$ for that in W44, when assuming a
nested model with two additional degrees of freedom.

\begin{table}[htdp]
\caption{Spectral parameters in the energy range of 60 MeV to 2 GeV
  for power-law (PL) and broken power-law (BPL) models. $TS = 2
  ln(\mathcal{L}_1/\mathcal{L}_2)$ is the test-statistic value. }
\begin{center}
\begin{tabular}{cccccc}
\hline 
Model & $K\  \rm  (cm^{2}\  s^{-1}\ MeV^{-1})$ & $\Gamma_1$ & $\Gamma_2$ & $\varepsilon_{\rm br}$ (MeV) & $TS$ \\
\hline 
\multicolumn{1}{l}{IC~443} & & & & & \\
PL & $11.7 \pm 0.2) \times 10^{-10}$ & $1.76\pm 0.02$ 
& $\cdots$ & $\cdots$  &  21651 \\
BPL & $(11.9 \pm 0.6) \times 10^{-10}$ & $0.57\pm 0.25$ 
& $1.95^{+0.02}_{-0.02}$ & $245^{+16}_{-15}$  &  22010\\
\multicolumn{1}{l}{W44} & & & & & \\
PL & $(13.0 \pm 0.4) \times 10^{-10}$ & $1.71\pm 0.03$ 
& $\cdots$ & $\cdots$  &  6920 \\
BPL & $(15.8 \pm 1.0) \times 10^{-10}$ & $0.07\pm 0.4$ 
& $2.08^{+0.03}_{-0.03}$ & $253^{+11}_{-11}$  &  7351 \\
\hline 
\end{tabular}
\end{center}
\label{table}
\end{table}

\begin{figure}[!t]
\centering
\includegraphics[width=\textwidth]{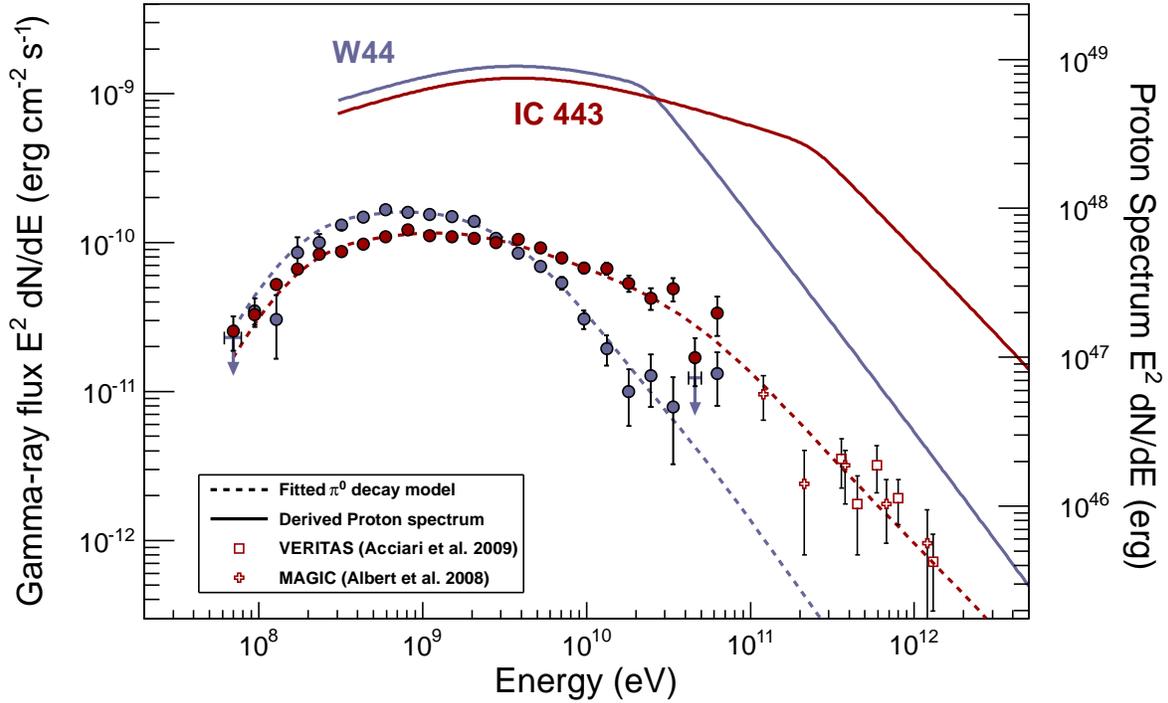}
\caption{ Proton and gamma-ray spectra determined for IC~443 and W44.
  Also shown are the broadband spectral flux points derived in this
  study, along with TeV spectral data points for IC~443 from MAGIC
  \cite{MAGIC_IC443} and VERITAS \cite{VERITAS_IC443}.  The curvature
  evident in the proton distribution at $\sim2$ GeV is a consequence
  of the display in energy space (rather than momentum space).
  Gamma-ray spectra from the protons were computed using the
  energy-dependent cross section parameterized by~\cite{Kamae06}. We
  took into account accelerated nuclei (heavier than protons) as well
  as nuclei in the target gas by applying an enhancement factor of
  1.85~\cite{Mori2009}. Note that models of the
  gamma-ray production via {\emph{pp}} interactions have some
  uncertainty. Relative to the model adopted here, an alternative
  model of ~\cite{Dermer86} predicts $\sim 30\%$ less photon flux
  near 70 MeV; the two models agree with each other to better
  than 15\% above 200 MeV.  The proton spectra assume average gas
  densities of $n = 20\ \rm cm^{-3}$ (IC~443) and $n = 100\ \rm
  cm^{-3}$ (W44) and distances of 1.5 kpc (IC~443) and 2.9 kpc
  (W44).}
\label{fig::proton}
\end{figure}

To determine whether the spectral shape could indeed be modeled
with accelerated protons, we fit the LAT spectral points with a
$\pi^0$-decay spectral model, which was numerically calculated from a
parameterized energy distribution of relativistic protons.  Following
previous studies \cite{Fermi_W44,Fermi_IC443}, the parent proton
spectrum as a function of momentum $p$ was parameterized by a smoothly
broken power law in the form of
\begin{eqnarray}
\frac{dN_p}{dp} \propto p^{-s_1} \left[1 + \left(
  \frac{p}{p_{\rm br}}\right)^{\frac{s_2-s_1}{\beta}} \right]^{-\beta}. 
\end{eqnarray}
Best-fit parameters were searched using $\chi^2$-fitting to the flux
points. 
The measured gamma-ray spectra, in particular the low-energy parts,
matched the $\pi^0$-decay model (Figure~\ref{fig::spec}). Parameters
for the underlying proton spectrum are $s_1 = 2.36\pm 0.02$, $s_2 =
3.1 \pm 0.1$, and $p_{\rm br} = 239 \pm 74~{\rm GeV}\, c^{-1}$ for
IC~443 and $s_1 = 2.36 \pm 0.05$, $s_2 = 3.5 \pm 0.3$, and $p_{\rm br}
= 22 \pm 8~{\rm GeV}\, c^{-1}$ for W44 (statistical errors only).  In
Figure~\ref{fig::proton} we show the energy distributions of the
high-energy protons derived from the gamma-ray fits. The break $p_{\rm
  br}$ is at higher energies and is unrelated to the low-energy
pion-decay bump seen in the gamma-ray spectrum.  If the interaction
between a cosmic-ray precursor (i.e., cosmic rays distributed in the
shock upstream on scales smaller than $\sim 0.1R$, where $R$ is the
SNR radius) and adjacent molecular clouds were responsible for the
bulk of the observed GeV gamma rays, one would expect a much harder
energy spectrum at low energies (i.e. a smaller value for the index
$s_1$), contrary to the \emph{Fermi} observations.  Presumably, cosmic
raysin the shock downstream produce the observed gamma rays; the first
index $s_1$ represents the shock-acceleration index with possible
effects due to energy-dependent propagation, and $p_{\rm br}$ may
indicate the momentum above which protons cannot be effectively
confined within the SNR shell.  Note that $p_{\rm br}$ results in the
high-energy break in the gamma-ray spectra at $\sim 20$~GeV and $\sim
2$~GeV for IC~443 and W44, respectively.

The $\pi^0$-decay gamma rays are likely emitted through interactions
between ``crushed cloud" gas and relativistic protons, both of which
are highly compressed by radiative shocks driven into molecular clouds
that are overtaken by the blast wave of the SNR \cite{Uchiyama10}.
Filamentary structures of synchrotron radiation seen in a
high-resolution radio continuum map of W44 \cite{Caste07} support this
picture.  High-energy particles in the ``crushed cloud" can be
explained by re-acceleration of the pre-existing galactic cosmic rays
\cite{Uchiyama10} and/or freshly accelerated particles that have
entered the dense region \cite{UchiyamaW44}.  The mass of the shocked
gas ($\sim 1\times 10^3 M_\odot$ and $\sim 5\times 10^3 M_\odot$ for
IC~443 and W44 respectively, where $M_\odot$ is the mass of the Sun)
is large enough to explain the observed gamma-ray luminosity.  Because
the ``crushed cloud" is geometrically thin, multi-GeV particles are
prone to escape from the dense gas, which may explain the break
$p_{\rm br}$.

Escaped CRs reaching the unshocked molecular clouds ahead of the SNR
shock can also produce $\pi^0$-decay gamma rays
\cite{Gabici,Ohira11}. Indeed, the gamma rays emitted by the escaped
CRs in the large molecular complex that surrounds W44 (total extent of
100 pc) have been identified with three close-by
sources\cite{UchiyamaW44}, which led us to remove them from the model
in the maximum likelihood analysis, as mentioned above. With this
treatment, the measured fluxes below 1 GeV contain small contributions
from the escaped CRs, but this does not affect our conclusions.  The
escaped CRs may significantly contribute to the measured TeV fluxes
from IC~443 \cite{MAGIC_IC443,VERITAS_IC443}. Emission models could be
more complicated. For example, the CR precursor with a scale of $\sim
0.1R$ at the highest energy could interact with the adjacent unshocked
molecular gas, producing hard gamma-ray emission. This effect is
expected to become important above the LAT energy range.

We should emphasize that radiation by relativistic electrons can not
as naturally account for the gamma-ray spectra of the
SNRs~\cite{SOM}. An inverse-Compton origin of the emission was not
plausible on energetic grounds\cite{Fermi_W51C}. The most important
seed photon population for scattering is the infrared radiation
produced locally by the SNR itself with an energy density of $\sim 1\
\rm eV\ cm^{-3}$, but this is not large enough to explain the observed
gamma-ray emission.
Unless we introduce in an ad hoc way an additional abrupt break in the
electron spectrum at 300~MeV~$c^{-1}$ (Fig.~\ref{fig::spec}
dash-dotted lines), the bremsstrahlung models do not fit the observed
gamma-ray spectra. If we assume that the same
electrons are responsible for the observed synchrotron radiation in
the radio band, a low-energy break is not expected to be very strong
in the radio spectrum and thus the existing data do not rule out this
scenario. The introduction of the low-energy break introduces
additional complexity and therefore a bremsstrahlung origin is not
preferred. Although most of the gamma-ray emission from these SNRs is
due to $\pi^0$-decay, electron bremsstrahlung may still contribute at
a lower level.  The \emph{Fermi} LAT data allow an electron-to-proton
ratio $K_{ep}$ of $\sim0.01$ or
less, where $K_{ep}$ is defined as the
ratio of $dN_e/dp$ and $dN_p/dp$ at $p=1$ GeV\,$c^{-1}$
(figs. \ref{fig::brems_ic443} and \ref{fig::brems_w44}).

Finding evidence for the acceleration of protons has long been a key
issue in attempts to elucidate the origin of cosmic rays. Our spectral
measurements down to 60 MeV enable the
identification of the $\pi^0$-decay feature, thus  providing direct evidence for the
acceleration of protons in SNRs.  The proton momentum distributions,
well-constrained by the observed gamma-ray spectra, are yet to be
understood in terms of acceleration and escape processes of
high-energy particles.

\bibliographystyle{Science}

\begin{scilastnote}
\item The {\emph Fermi} LAT Collaboration acknowledges support from a
  number of agencies and institutes for both development and the
  operation of the LAT as well as scientific data analysis. These
  include NASA and the U.S. Department of Energy (United States); CEA/Irfu and IN2P3/CNRS
  (France); ASI and INFN (Italy); MEXT, KEK, and JAXA (Japan); and
  the K.~A.~Wallenberg Foundation, the Swedish Research Council and
  the National Space Board (Sweden). Additional support from INAF in
  Italy and CNES in France for science analysis during the operations
  phase is also gratefully acknowledged. Fermi LAT data are available
  from the Fermi Science Support Center
  (http://fermi.gsfc.nasa.gov/ssc).
\end{scilastnote}

\clearpage

\clearpage

\clearpage

\clearpage

\makeatletter 
\setcounter{figure}{0}
\renewcommand{\thefigure}{S\@arabic\c@figure} 
\makeatother 

\section*{Supplementary Material} 

\subsection*{Analysis of the LAT data}

In its normal mode, the LAT scans the whole sky every 3 hours (two
orbits). In this analysis, data taken in the sky-survey mode starting
from the beginning of scientific operation on 2008 August 4 to 2012
July 16 were used. These data were analyzed using the LAT {\it Science
  Tools} package (v9r29), which is available from the {\emph{Fermi}} Science
Support Center. The analysis used the {\emph{P7v6}} version of the
instrument response functions which take into account accidental
coincidence effects in the detectors. Only events passing the
``Source'' class cuts are used in the analysis, and events coming from
zenith angles $> 100^{\circ}$ were discarded to reduce the
contribution from the emission in the Earth's upper atmosphere. To
further reduce these Earth-emitted gamma rays, intervals when the
Earth was in the field of view were excluded, specifically when the
rocking angle of the LAT was greater than $52^{\circ}$ or when parts
of the region-of-interest (ROI) were observed at zenith angles
$>100^{\circ}$. All gamma rays with energies $>60$ MeV within a
$20^{\circ} \times 20^{\circ}$ region around the nominal positions of
the sources were used.

During the broad-band fit of the source of interest, all 2FGL sources
within the field of view were part of the likelihood model with the
spectral models as adopted in the 2FGL catalog~\cite{2FGL}. Because
the 2FGL catalog corresponds to 2 years of data, while the dataset
used in the study comprises 4 years, we searched for additional
(faint) sources within the ROI by constructing maps of residual
significance after subtracting all known sources in the field. For the
W44 region three additional sources were identified at positions
($\alpha_{\rm J2000}, \delta_{\rm J2000}$)= $(283.92^{\circ},
2.79^{\circ})$ with a TS value above 60~MeV of $137$,
$(285.25^{\circ}, 7.11^{\circ})$ with a TS value of $98$, and
$(290.90^{\circ}, 6.62^{\circ})$ with a TS value of $52$, where
$\alpha_{\rm J2000}$ and $\delta_{\rm J2000}$ are right ascension and
declination for the epoch 2000.  In the IC~443 ROI one additional
source was identified at ($\alpha_{\rm J2000}, \delta_{\rm J2000}$)=
$(100.17^{\circ}, 17.80^{\circ})$ with a TS value of 114. None of
these sources affect the spectral parameters fit for the sources under
study but they were included for completeness and to make the residual
maps flat. To account for correlations between close-by sources, the
normalizations of close-by sources (shown as crosses in
Figure~\ref{fig::cmap}) and of the Galactic and isotropic diffuse
emission were left free in the broad-band and in the bin-by-bin
fitting procedure. In the case of IC~443, the Crab was modeled with
three components for which the normalization was left free
following~\cite{CrabPaperII}: one for the Crab pulsar, and two for the
synchrotron and for the inverse Compton part of the Crab Nebula
respectively. The maximum likelihood estimation for spatially and
spectrally binned data is performed inside a square region of
$20^{\circ}\times20^{\circ}$ centered on each SNR (corresponding to
the regions shown in Figure~\ref{fig::cmap}).  The spatial model for
both SNRs was taken as a uniform disk with an angular radius of
$21^{\prime}$\cite{ExtendedFermi}.  The exact choice of the template
has only a small impact on the measurement of the spectrum at low
energies, due to the large point-spread function of the instrument
(the 68\% containment radius is $6^{\circ}$ at 100 MeV), while it
affects the spectrum above 1 GeV at the $< 30\%$ level
\cite{Fermi_W51C}.

\subsection*{Systematic errors} 

For regions near the Galactic equator, the dominant systematic error
at energies below $\sim 1$~GeV arises from the uncertainty in the
model of the Galactic diffuse emission.  To test for this effect on
the emission assigned to IC~443 and W44, a range of alternative
diffuse models calculated by GALPROP~\cite{StrongMoskalenko98} and
then adjusted for agreement with the observed diffuse gamma-ray
emission within the ROI was used. The GALPROP models were chosen to
represent a broad range of the parameters scanned
in~\cite{FermiDiffuse2}. They were adjusted for overall consistency
with the all-sky gamma-ray data in a likelihood fit. When fitting the
alternative diffuse models in this analysis, the overall normalization
was varied. Eight models were tested: the parameters varied among the
models are the radial distribution of the cosmic-ray sources (SNR-like
or pulsar-like), the size of the cosmic-ray halo (4 kpc or 10 kpc)
and the spin temperature of atomic hydrogen (150 K or optically
thin). It is not a priori obvious which of these (or the standard)
diffuse models are the best descriptions of the data for the particular
ROIs under study, but the range of diffuse models is intended to test
systematics related to uncertainties in the model for the Galactic
diffuse emission. The gray bands in Figure~\ref{fig::spec} show the
systematic error due to uncertainties in the Galactic diffuse
emission, derived as the envelope of the broad-band fits for the
different models.

\begin{figure}[ht]
\centering
\includegraphics[width=\textwidth]{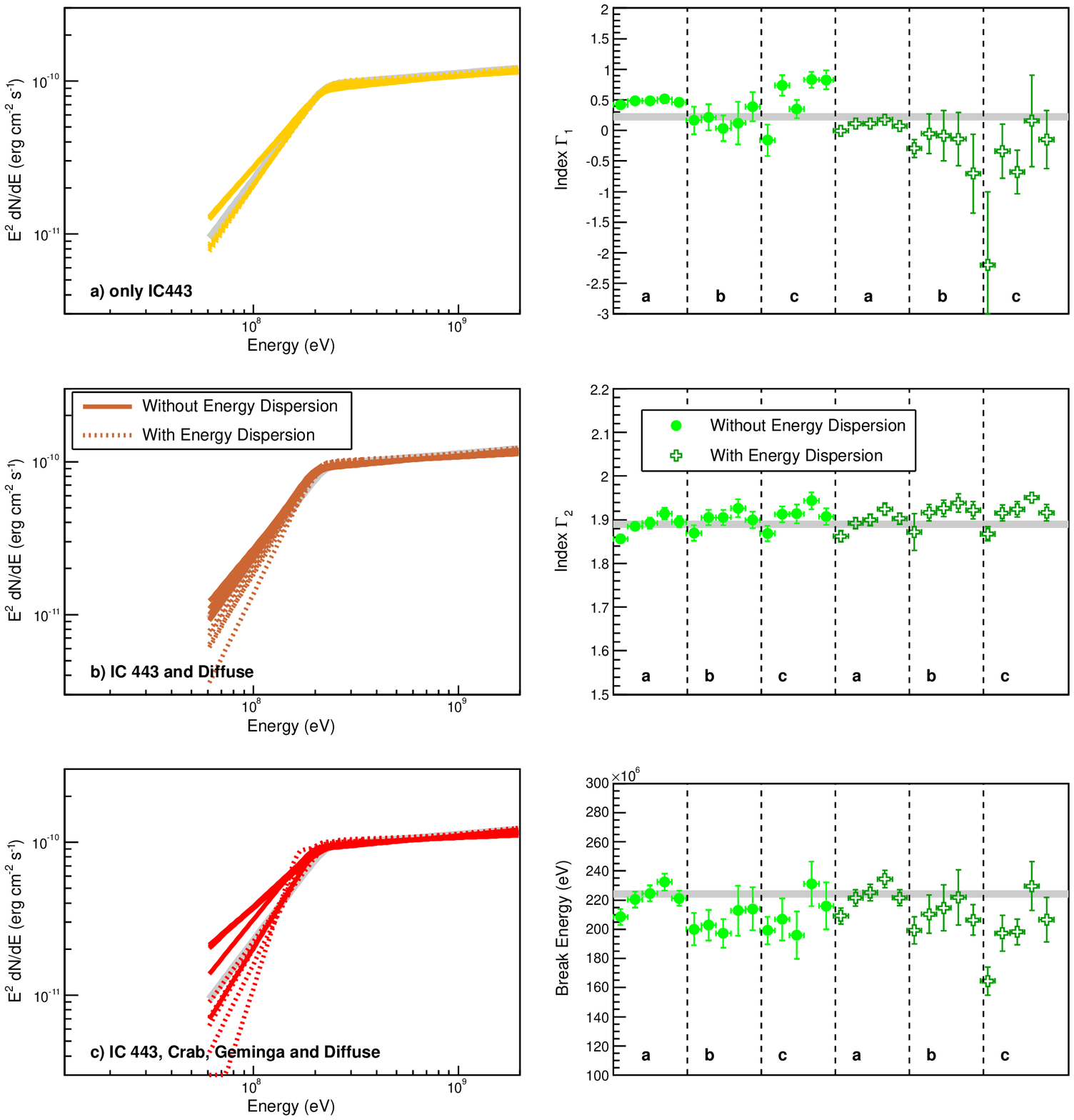}
\caption{\small The effect of ignoring or taking the energy dispersion into
  account in the fitting of simulated data (with energy dispersion) is
  demonstrated here. For each case 5 realizations of the IC~443 ROI
  were simulated. The results from fitting with the energy dispersion
  ignored are shown as solid lines, the results from fitting with the
  energy dispersion taken into account are shown as dotted lines. The
  gray lines show the input spectrum. The left panel shows a
  simulation of IC~443 only, the middle panel a simulation with IC~443
  and the Galactic and isotropic diffuse and the right panel a
  simulation with IC~443, Crab and Geminga and the Galactic and
  isotropic diffuse emission. The lower row shows the deviations from
  the input spectrum in the reconstructed values of $\Gamma_1$,
  $\Gamma_2$ and the break energy $E_{\mathrm{br}}$. The solid points
  show the cases in which the energy dispersion was ignored, the open
  crosses the cases in which the energy dispersion was switched on. As
  can be seen, the main effect is on $\Gamma_1$ as expected which is
  $\sim 1$ steeper if energy dispersion is taken into account.}
\label{fig::edisp}
\end{figure}

An additional important source of systematic errors that needs to be
considered when analyzing LAT data at low energies is the effect of
energy dispersion. In the standard analysis of LAT data, energy
dispersion is typically neglected. As long as the LAT effective
collection area is relatively independent of energy (above $\sim200$
MeV) the bias on the flux in an energy bin that arises from
misreconstructing the energies of the events is relatively mild (of
the order 1-2\%). However, the combination of energy dispersion and
the rapidly changing effective area below 100 MeV for the instrument
response functions used in this study (P7SOURCE\_V6) does result in
biased measurements of flux and spectral index of the source under
study. In order to quantify the effects of energy dispersion,
simulations of the IC~443 region were performed taking into account
the observation conditions of these sources in the real dataset and
simulating all nearby sources in the region. These simulations include
the effect of energy dispersion. The analysis performed on these
simulations was exactly the same as performed on the real data. For
the study presented here two effects are important: the low-energy
index in the broad-band fit ($\Gamma_1$) will be less steeply falling
when going to lower energies and the flux points below $\sim 150$ MeV
will be shifted upwards when ignoring energy dispersion. The
simulations can be used to quantify the effect when ignoring energy
dispersion. For the spectrum of IC 443, $\Gamma_1$ is typically less
steeply falling by $\sim 1$ (see Figure~\ref{fig::edisp}) when
ignoring the effect. The flux points are higher by the following
fractions ($60-81~\mathrm{MeV}: 39\%, 81-110~\mathrm{MeV}: 23\%,
110-150~\mathrm{MeV}: 12\%$). For W44 we expect corrections of the
same magnitude. At higher energies these factors become smaller than
5\%. These studies demonstrate that taking energy dispersion in the
analysis into account results in the correct reconstructed values. The
magnitude of the bias of the index depends on the spectrum of the
Galactic diffuse emission model at energies much lower than the 60 MeV
limit for the present analysis.  Unlike the simulations just described
we do not have a perfect model for the diffuse emission, and so even
with energy dispersion included in the calculations, $\Gamma_1$ is
difficult to determine accurately, although for any plausible model
the sign of the bias in $\Gamma_1$ from neglecting energy dispersion
is always the same.  In order to not overestimate the magnitude of the
spectral break we have decided to neglect energy dispersion in the
spectral analysis for this paper. This decision will make the measured
low-energy break less pronounced but has little impact on the fitted
spectrum for the cosmic-ray protons responsible for the gamma rays,
since it just affects the exact shape of the pion bump.

\subsection*{Leptonic models}
In addition to $\pi^0$-decay emission, bremsstrahlung and inverse
Compton scattering from electrons are possible gamma-ray emission
mechanisms for SNRs.  Here we consider the cases in which
the leptonic processes are dominant in the energy range covered by
\emph{Fermi}-LAT.

\begin{figure}[h]
\centering
\includegraphics[width=\textwidth]{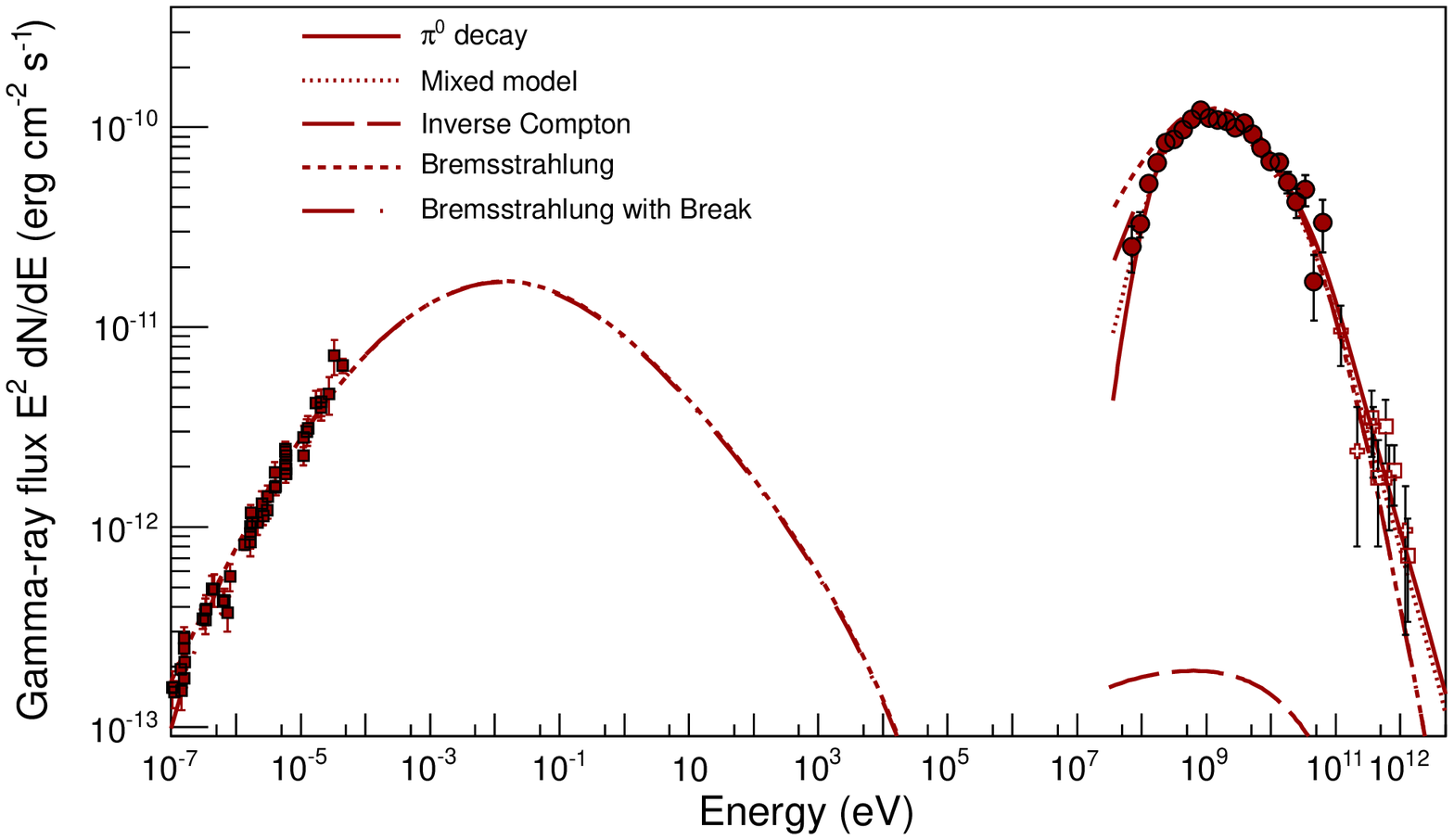}
\caption{ \small The spectral energy distribution of
    IC~443 with a model in which electron bremsstrahlung is the
    dominant radiation process in the gamma-ray band (dashed line and
    dash-dotted as in Figure 2). In the radio band the dominant
    radiation process is synchrotron emission.  The parameters used in
    the calculation for the dashed curve are $s_1 = 1.72$, $s_2 =
    3.2$, $p_{\rm br} = 10~{\rm GeV}~c^{-1}$, $B = 50~\mu {\rm G}$, $n
    = 300~{\rm cm}^{-3}$, and $W_e = 5 \times 10^{47}~{\rm erg}$ ($>
    1~{\rm GeV}~c^{-1}$). The dash-dotted curve indicates the same
    model but with an abrupt low-energy break in the electron
    spectrum at 300~MeV$c^{-1}$. For comparison the best-fit
    pion-decay model is shown as a solid line. The dotted line shows a
    combined bremsstrahlung and pion-decay model in which $K_{ep}=
    0.01$ to demonstrate that such a model is consistent with the
    data. Radio data are taken from \cite{Caste11}}.
\label{fig::brems_ic443}
\end{figure}

As already discussed in the main text, it is difficult to explain the
large gamma-ray luminosity of W44 and IC~443 with inverse Compton
scattering (see e.g. dashed curve in Figure~\ref{fig::brems_ic443}).
Considering typical interstellar radiation fields and infrared photons
produced locally by the SNR as target photons, the total kinetic
energy of electrons is required to be $\sim 10^{51}~{\rm erg}$, which
means almost 100\% of the kinetic energy released by a supernova
explosion should be consumed for electron acceleration.  In addition,
the measurement of a sharply falling spectrum below $\sim 200$ MeV is
inconsistent with the radio spectrum.

\begin{figure}[h]
\centering
\includegraphics[width=\textwidth]{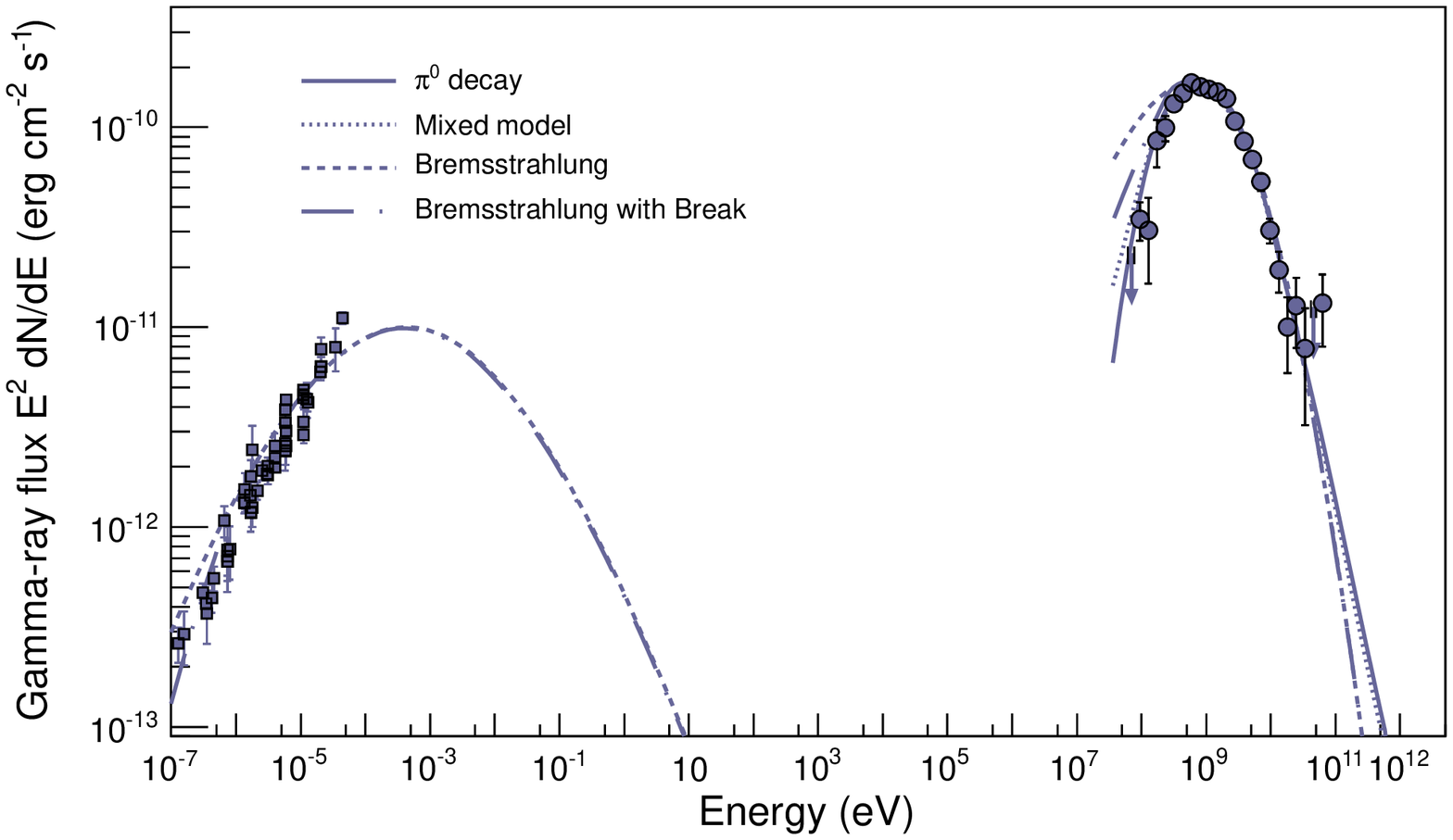}
\caption{ \small The same as Figure~\ref{fig::brems_ic443} but for
  W44. The parameters are $s_1 = 1.74$, $s_2 = 3.7$, $p_{\rm br} =
  10~{\rm GeV}~c^{-1}$, $B = 90~\mu {\rm G}$, $n = 650~{\rm cm}^{-3}$,
  and $W_e = 6 \times 10^{47}~{\rm erg}$ ($> 1~{\rm
    GeV}~c^{-1}$). Radio data are taken from  \cite{Caste07}.}
\label{fig::brems_w44}
\end{figure}

We thus consider models in which electron bremsstrahlung is dominant
in the LAT energy band. We assume for the electrons a smoothly broken
power law, similar to what was used for the proton spectrum.  The
radio data of SNRs IC~443 and W44 strongly constrain the electron
indices below the break ($s_1$).  On the other hand, the index above
the break ($s_2$) and the location of the break $p_{\rm br}$ can be
determined by fitting the gamma-ray spectral shapes.  The total number
of electrons, magnetic field strength, and ambient gas density are
then chosen to simultaneously fit the synchrotron radio and the
gamma-ray bremsstrahlung flux.  The magnetic field cannot be too low
otherwise a break in the synchrotron spectrum corresponding to
$p_{br}$ appears in the radio band where pure power-law-type spectra
are observed.  Figures \ref{fig::brems_ic443} and \ref{fig::brems_w44}
show the models for IC~443 and W44, respectively (dashed lines).  An
important discrepancy can be found in the energy range below $<
200$~MeV, where we found low-energy breaks in the \emph{Fermi}-LAT
data.

In order to match the \emph{Fermi}-LAT data, the electron spectra need
to have additional low-energy breaks.  In Figures
\ref{fig::brems_ic443} and \ref{fig::brems_w44}, we plot
bremsstrahlung models with the low-energy breaks added to the electron
spectra (dash-dotted lines).  To make the bremsstrahlung spectra below
$\sim 200$~MeV as hard as possible, we have applied abrupt breaks at
300~MeV~$c^{-1}$ in the electron spectra for both W44 and IC~443.
Even with this extreme assumption, the model curves do not become as
hard as the \emph{Fermi}-LAT spectra below $\sim 200$~MeV as seen in
Figure~\ref{fig::spec}, although the model and data are marginally
consistent considering the systematic errors.

Our results indicate most of the GeV gamma-ray emission is of hadronic
origin, i.e., $\pi^0$-decay gamma rays (solid lines).  The observed
synchrotron radio emission still constrains the accelerated electrons
in the SNRs.  As shown in Figures~\ref{fig::brems_ic443} and
\ref{fig::brems_w44}, the radio spectra, measured at frequencies from
several 10~MHz up to 10~GHz, have a power-law shape with spectral
indices of $\alpha =0.36$ and 0.37 for IC~443 and W44, respectively.
This means that the radio-emitting electrons have a power-law index of
$s_1 = 1.72$ and 1.74 for IC~443 and W44, respectively.  The electron
indices are smaller than the proton indices of $s_1 = 2.3$--2.4
determined from the gamma-ray spectra.  It should be noted that the
radio-emitting electrons have lower momentum compared with the
GeV-emitting protons in the crushed cloud scenario of
~\cite{Uchiyama10, UchiyamaW44}, and therefore different indices would
be explained by introducing another spectral break.  Moreover,
momentum spectra of electrons and protons do not have the same shape
at low energies if re-acceleration of the Galactic cosmic rays is
responsible for the observed radio and gamma-ray emission
\cite{Uchiyama10}.  Given that middle-aged, partially radiative SNRs
with molecular cloud interactions are complex systems, there is
another possibility that the radio-emitting electrons and GeV-emitting
protons have different origins, particularly in the case of IC~443.
Strong radio emission in IC~443 comes from the northeastern part of
the remnant where the shock is propagating in atomic clouds, while the
gamma-ray peaks are located near the interacting molecular clouds.

\subsection*{Energetics}

As discussed in the main text, the dominant sites of gamma-ray
emission are likely to be the shocked molecular clouds.  The mass of
the shocked gas is estimated as $M_{\rm shocked} \sim 1\times 10^3
M_\odot$ and $\sim 5\times 10^3 M_\odot$ for IC~443 and W44,
respectively. To explain the gamma-ray luminosity, the CR energy
density in the clouds should be $\sim 400\ \rm eV\ cm^{-3}$, much
larger than that of the Galactic CRs.  Therefore, adiabatic
compression of the Galactic CRs alone cannot explain the required
energy density; we need re-acceleration of the pre-existing Galactic
CRs \cite{Uchiyama10} and/or freshly accelerated particles that have
entered the dense region \cite{UchiyamaW44}.  The total energy content
in protons with $p \geq 0.8\ {\rm GeV}\, c^{-1}$ (above the pion
production threshold) amounts to $W_{\rm CR} \simeq 4 \times 10^{49}
(n/20~{\rm cm}^{-3})^{-1}~{\rm erg}$ and $4 \times 10^{49} (n/100~{\rm
  cm}^{-3})^{-1}~{\rm erg}$ for IC~443 and for W44, respectively.
Here $n$ denotes the effective gas number density, given by $n =
M_{\rm shocked}/V_{\rm shell}$ where $V_{\rm shell}$ is the volume of
the SNR shell that contains shock-accelerated CR particles.  We assume
that the measured gamma-ray emission is produced mainly by CRs with
spectra given by Eq.~(1) interacting with the shocked clouds inside
SNRs. Also, the CR density is assumed to be uniform in the SNR shell,
which contains the shocked clouds.  Comparing estimates of the
explosion energies ($W_{\rm SN} \sim 1\times10^{51}$ ergs for
IC~443\cite{Troja2008} and $\sim 5\times10^{51}$ ergs for
W44\cite{Reach05}) with $W_{\rm CR}$, we obtain $W_{\rm CR}/W_{\rm
  SN}$ of the order of $1\mbox{--}10\%$ in these objects at their
current ages. The ratio strongly depends on the (rather uncertain)
assumed density at the location of the gamma-ray production.

\clearpage 

\end{document}